\documentclass[preprint,aps,titlepage,nofootinbib,showkeys,showpacs]{revtex4}

\usepackage{epsfig}
\usepackage{bm}
\usepackage{amsmath}

\newcommand{\x}{{\vec x}}
\newcommand{\y}{{\vec y}}
\newcommand{\Tr}{{\rm Tr}\,}
\renewcommand{\d}{\partial}
\newcommand{\<}{\langle}
\renewcommand{\>}{\rangle}

\newcommand{\be}{\begin{eqnarray}}

\newcommand{\ee}{\end{eqnarray}}
\newcommand{\ben}{\begin{eqnarray*}}
\newcommand{\een}{\end{eqnarray*}}

\def\eq#1{{Eq.~(\ref{#1})}}

\begin{document}

\preprint{NT@UW--02--036}
\preprint{INT--PUB--02--57}
\vspace*{1cm}
\setcounter{footnote}{1}

\title{Critical Temperature of the Deconfining Phase Transition in~\\~~\\
(2+1)d Georgi-Glashow Model}

\author{\vspace*{1cm}Yu.~V.~Kovchegov}\email{yuri@phys.washington.edu}

\affiliation{Department of Physics, University of Washington,
Seattle, Washington 98195-1560}

\author{D.~T.~Son}\email{son@phys.washington.edu}

\affiliation{Institute for Nuclear Theory, University of Washington,
Seattle, Washington 98195-1550\vspace*{1cm}}

\begin{abstract}
We find the temperature of the phase transition in the (2+1)d
Georgi-Glashow model.  The critical temperature is shown to depend on
the gauge coupling and on the ratio of Higgs and gauge boson masses.
In the BPS limit of light Higgs the previous result by Dunne, Kogan,
Kovner, and Tekin is reproduced. \vspace*{7cm}
\end{abstract}


\vspace*{2cm}
\maketitle

\section{Introduction}

Confining gauge theories generally become deconfined at high
temperatures.  The deconfining phase transition in QCD is the subject
of intensive experimental and lattice studies.  While some general
features of this phase transition can be extracted from symmetry
arguments~\cite{SvetitskyYaffe}, the details cannot be studied
analytically due to strong coupling effects resulting from the
non-perturbative nature of the phase transition.  Therefore it is
instructive to investigate theories where deconfinement happens in the
weak-coupling regime and hence can be studied quantitatively.  One
example of such theory, the (2+1)-dimensional SU(2) Georgi-Glashow
model, is considered in this paper. The model consists of a SU(2)
gauge theory coupled to a scalar (Higgs) field in the adjoint
representation with a non-zero VEV $v$.  By studying this model one
may hope to find certain features of the deconfining phase transitions
which this model shares with QCD. At zero temperature the
(2+1)-dimensional SU(N) Georgi-Glashow model could be related through
dimensional reduction to high temperature SU(N) gauge theory in four
dimensions \cite{lattice,Nadkarni}.

It has been shown by Polyakov~\cite{Polyakov} that 't Hooft--Polyakov
monopoles \cite{thp} [which are instantons in (2+1)d] give rise to the
area law for Wilson loops, i.e., confinement.  From the symmetry
properties of the Polyakov loops~\cite{SvetitskyYaffe} it follows that
the Georgi-Glashow model must possess a sharp deconfining phase
transition at some temperature. The first analytical attempt to find
the temperature of the phase transition in the limit of the small
gauge coupling $g \ll v$ was made by Agasyan and Zarembo~\cite{AZ}.
Following \cite{Polyakov} they argued that at low temperatures ($T \ll
M_W$) the theory is equivalent to a 2-dimensional gas of monopoles
interacting with each other via (almost massless) Coulomb
photons. (Remember that the original SU(2) symmetry is broken to
U(1).) The phase transition in the monopole gas would occur through
the binding of monopoles and anti-monopoles into pairs similar to
Berezinsky-Kosterlitz-Thouless (BKT) phase transition \cite{bkt}.
From this picture they found the critical temperature to be
$T_c=g^2/(2\pi)$ \cite{AZ}, where $g$ is the dimensionful gauge
coupling.  A criticism of this treatment was subsequently raised by
Dunne, Kogan, Kovner, and Tekin~\cite{DKKT}.  They argued that the
correct description of the low temperature limit of the Georgi-Glashow
model should include $W$ bosons interacting with each other and with
the monopoles by exchanging photons. The model thus reduced to a
two-dimensional gas of point charges ($W$'s) and monopoles.  The
presence of thermally excited $W$ bosons decreased the critical
temperature by a factor of two yielding $T_c=g^2/(4\pi)$ \cite{DKKT}.
The authors of \cite{DKKT} also showed that the $W$'s are required for
the phase transition to be of the $Z_2$ universality class, in
agreement with symmetry arguments~\cite{SvetitskyYaffe}.

In this paper we re-examine the calculation of the critical
temperature.  We shall show that it actually depends on both the gauge
coupling $g$ and the ratio between the Higgs mass $m_H$ and the gauge
boson mass $m_W$.  Specifically, our result reads
\begin{equation}
  T_c = \frac{g^2}{4\pi} \frac{\epsilon+2}{2\epsilon+1},
  \label{Tc}
\end{equation}
where $\epsilon=\epsilon(m_H/m_W)$ is a function of the ratio of Higgs
and $W$ masses $M_H$ and $M_W$ defining the monopole action
\cite{bps,kz}
\begin{equation}\label{Sinst}
  S_0 = \frac{4\pi v}g \epsilon\left(\frac{m_H}{m_W}\right).
\end{equation}
When $m_H\ll m_W$, which is the limit where the
Bogomolny-Prasad-Sommerfield (BPS) bound \cite{bps} on the monopole
action is saturated, $\epsilon (0) = 1$ and $T_c$ in Eq.~(\ref{Tc})
coincides with the value found in Ref.~\cite{DKKT}.  Outside this
limit, our result disagrees with both Refs.~\cite{DKKT} and \cite{AZ}.

Our derivation of Eq.~(\ref{Tc}) is based on what we believe to be the
correct interpretation of the renormalization group approach initiated
in Ref.~\cite{DKKT}.  To make the paper self-contained, we will
provide alternative derivations for many facts already known in
Refs.~\cite{AZ,DKKT}.  Readers already familiar with
Refs.~\cite{AZ,DKKT} can jump directly to Sec.~\ref{sec:RG}, which is
the core of our paper where Eq.~(\ref{Tc}) is derived.

The paper is organized as follows.  In Sec.~\ref{sec:model} we
formulate the Georgi-Glashow model, mostly in order to introduce
notations.  In Sec.~\ref{sec:Coulomb} we rewrite the partition
function in term of a 2d Coulomb gas of electric and magnetic charges.
In Sec.~\ref{sec:effective} we show that the Coulomb gas, in its turn,
can be recast into a field theory.  We give two equivalent
representations of this theory in terms of bosonic and fermionic
variables. Our bosonic field theory is different from the one used in
\cite{DKKT}.  Sec.~\ref{sec:RG} is devoted to the renormalization
group analysis of the field theory where we determine the value of the
critical temperature.  Comparison with earlier works and a simple
physical argument rederiving \eq{Tc} are given in
Sec.~\ref{sec:concl}.  In the Appendix we review the free massless
scalar in two dimensions and show how a bosonic theory for the gas of
electric and magnetic charges is constructed.

\section{The model}
\label{sec:model}

In this paper we shall work mostly in Euclidean space.  The
Georgi-Glashow model is a SU(2) gauge theory where the gauge symmetry
is partially broken by a triplet scalar,
\begin{equation}
  L = \frac12 (D_\mu\phi^a)^2 + \lambda(|\phi|^2-v^2)^2 
  + \frac14 F_{\mu\nu}^2
\end{equation}
In the unitary gauge where $\phi^1=\phi^2=0$, the photon field $A^3$
is perturbatively massless, while $W^\pm = A^1\pm i A^2$ are massive
gauge bosons with the mass $m_W=g v$ and are charged with respect to
the photon field.  The theory contains two dimensionless parameters:
$g/v$, which is the expansion parameter of the perturbation theory,
and $m_H/m_W\sim\sqrt\lambda/v$.  We shall assume $g/v\ll1$ and
$m_H/m_W$ parametrically of order one, that is $m_W \sim m_H \gg g^2$.

The model contains a classical solution --- the 't Hooft--Polyakov
``monopole'', which is really an instanton in (2+1)d. (We shall use
the terms ``monopole'' and ``instanton'' interchangeably.)  At large
distances, in the unitary gauge, the U(1) field of the instanton is
\begin{equation}
  F^3_{\mu\nu} = \frac e{4\pi}\epsilon_{\mu\nu\lambda} 
  \frac{x_\lambda}{x^2}\,,
\end{equation}
where $e$ is the magnetic charge of the monopole,
\begin{equation}\label{e4pig}
  e = \frac{4\pi}g\,.
\end{equation}
In our case $g e$ is twice larger than the Dirac quantum $2\pi$.  The
monopole action is given by Eq.~(\ref{Sinst}) and is large at weak
coupling $g/v\ll1$; correspondingly the density of instantons is
exponentially small, $\sim e^{-S_0}$.  Due to the contributions of
these instantons to the partition function, photon acquires a mass
$m_\gamma$ which is exponentially suppressed, $m_\gamma \sim
e^{-S_0/2}$.

In the Euclidean formalism, one puts the field theory at a finite
temperature $T$ by making the Euclidean time direction periodic with
the period $\beta=1/T$.  It turns out that the interesting physics of
deconfinement occurs at temperature of order $g^2$, which is much
smaller than $W$ and Higgs masses but is much larger than the photon
mass $m_\gamma$.  We shall concentrate on the behavior of the theory
at such temperatures.

\section{Coulomb gas representation}
\label{sec:Coulomb}

Our goal is to rewrite the partition function of the Georgi-Glashow
model at finite temperature $T$ such that $m_\gamma \ll T \ll m_W \sim
m_H$ in terms of an effective two-dimensional theory.  Extending
Ref.~\cite{Polyakov}, our first step is to rewrite the partition
function in terms of a two-dimensional Coulomb gas of monopoles and
$W$ bosons,
\begin{equation}\label{z1}
  Z = \sum_{M,N=0}^\infty \frac1{M!N!} \sum_{e_i,g_j}\int\!
  \prod_{i=1}^M d^2x_i \prod_{j=1}^N d^2y_j\, 
  e^{-S(\vec x_i, e_i; \vec y_j, g_j)}
\end{equation}
where $S(\vec x_i, e_i; \vec y_j, g_j)$ is the effective action of a
system of $M$ monopoles located at the (2d) positions $\vec x_i$ with
magnetic charges $e_i$, and $N$ $W$-bosons located at positions $\vec
y_j$ with electric charges $g_j$.  The sum over $e_i,g_j$ in \eq{z1}
goes over different signs of the $W$-boson and monopole charges. The
effective action is the sum of three contributions,
\begin{equation}\label{S=S+S+S}
  S(\vec x_i, e_i; \vec y_j, g_j) = S_{\rm mon}(\vec x_i, e_i) +
  S_W(\vec y_j, g_j) + S_{\rm int}(\vec x_i, e_i;\vec y_j, g_j)
\end{equation}
where $S_{\rm mon}$ is the part of the action coming solely from the
monopoles, $S_W$ is coming solely from the $W$'s, and $S_{\rm int}$
represents the interaction between the monopoles and the $W$'s.

The explicit form of each term in Eq.~(\ref{S=S+S+S}) can be found
from physically intuitive arguments.  For $S_{\rm mon}$, we notice
that the monopole gas is effectively two-dimensional because our
assumption about the temperature $T\sim g^2 \gg m_\gamma \sim
e^{-S_0/2}$ makes the size of the compactified time direction ($1/T$)
much smaller than the average distance between the monopoles ($\sim
e^{S_0/2}$).  Thus, the monopoles interact via the 2d Coulomb
potential~\cite{AZ}, and
\begin{equation}\label{Smon}
  S_{\rm mon}(\vec x_i, e_i) = M S_0 
  - \frac{e^2T}{2\pi}\sum_{ij} e_i e_j\ln (T|\vec x_i-\vec x_j|)\,,\qquad
  \sum_i e_i = 0\,.
\end{equation}
Here $e_i=\pm1$ are the signs of the monopole charges, which are
measured in the units of $e$ defined in Eq.~(\ref{e4pig}).  The action
in \eq{Smon} integrated over all monopoles spatial positions is finite
only when the total magnetic charge vanishes; otherwise it is
logarithmically divergent.  The reason why $T$ appears in the argument
of the logarithms in Eq.~(\ref{Smon}) is that $T^{-1}$ is the smallest
distance where the inter-monopole potential is logarithmic.

Turning to $W$ bosons, we notice that since $T\ll M_W$ the $W$'s are
almost static.  As they carry U(1) electric charges they also form a
Coulomb gas.  We can visualize the worldlines of the $W$ bosons as
small circles wrapping around the Euclidean time direction; at
distances large compared to $\beta=1/T$, the ensemble of such circles
are indistinguishable from a gas of point-like objects.  The action
$S_W$ is simply the exponent in the Boltzmann suppression factor,
i.e., $E_W/T$ where $E_W$ is the energy of the $W$ configuration.
Therefore for almost static $W$'s
\begin{equation}
  S_W(\vec y_j, g_j) = N \frac{m_W}T 
  - \frac{g^2}{2\pi T}\sum_{ij} g_i g_j \ln(T|\vec y_i-\vec y_j|)\,,\qquad
  g_i=\pm 1\,,\qquad
  \sum_i g_i = 0\,.
\end{equation}
The first term in the right hand side comes from the static energy,
and the second term from the interaction between the $W$'s.  Although
the contribution of $W$ bosons to the free energy is exponentially
suppressed by $e^{-m_W/T}$, they can still destroy confinement, since
confinement in this model is due to instantons and the density of
instantons is also exponentially suppressed (\cite{DKKT}, also see
below).

The origin of the interaction term $S_{\rm int}$ is less trivial: it
is due to the complex phases acquired by the $W$ bosons along their
trajectories (which are closed time loops) in the U(1) field created
by the monopoles.  This purely imaginary contribution to the effective
action of the Coulomb gas has been explicitly computed in
ref.~\cite{AZ} to be
\begin{equation}\label{Sint}
  -S_{\rm int} = 
  2i \sum_{i=1}^N\sum_{j=1}^M e_ig_j \theta(\vec x_i-\vec y_j)
\end{equation}
where $\theta(\vec x_i-\vec y_j)$ is the angle between the vector
connecting the monopole at $\vec x_i$ and the $W$ boson at $\vec y_j$
and a chosen spatial direction (say, the $x_2$ axis).  The sum in
Eq.~(\ref{Sint}) does not depend on the choice of the reference
direction if the gas is electrically or magnetically neutral, or both.
Thus it is always well-defined when $S_{\rm mon}$ and $S_W$ are
finite.

\begin{figure}[h]
\begin{center}
\def\epsfsize #1#2{0.6#1}
\epsffile{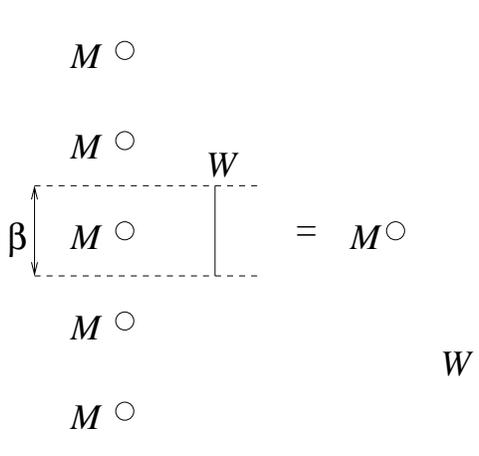}
\end{center}
\caption{The phase obtained by a $W$ in the field of periodic
instantons (see text).}
\label{fig:period}
\end{figure}

Equation~(\ref{Sint}) can be derived in a simple, almost geometrical,
way.  Let us consider a simple case of one monopole and and a pair of
$W^+$ and $W^-$ bosons.  First we ``unwind'' the time direction, so
that we have $W$ bosons move along finite time intervals in the field
of an infinite number of monopoles, positioned periodically along the
time axis.  It is obvious from Fig.~\ref{fig:period} that the phase
obtained by a $W$ boson in this case is the same as when it moves
along an infinite temporal trajectory in the field of a single
monopole.

\begin{figure}
\begin{center}
\def\epsfsize #1#2{0.4#1} 
\epsffile{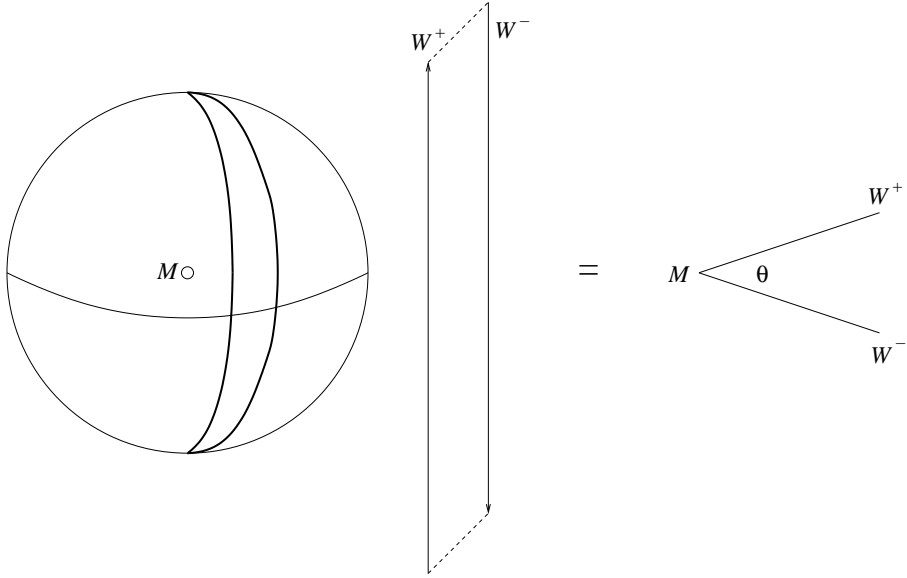}
\end{center}
\caption{Geometrical interpretation of the phase obtained by a pair of
$W^+$ and $W^-$ boson in the field of a monopole.}
\label{fig:flux}
\end{figure}

Now the total phase obtained by a pair of $W^+$ and $W^-$ bosons is
proportional to the magnetic flux passing through the two-dimensional
surface stretched between the worldlines of the two bosons (see
Fig.~\ref{fig:flux}).  If we surround the monopole by a sphere, then
the flux is proportional to the area between the two meridians as
shown in Fig.~\ref{fig:flux}.  Projected to the equator plane, one
sees that flux is indeed proportional to the angle between the line
connecting the monopoles to the $W^+$ and $W^-$ bosons, in agreement
with Eq.~(\ref{Sint}).  The coefficient 2 in Eq.~(\ref{Sint}) is due
to the fact that the U(1) charge of $W$ bosons is twice the elementary
value.

Gathering all terms, we arrive to the Coulomb-gas representation of
the partition function of the Georgi-Glashow models at temperatures large
compared to $m_\gamma$, but small compared to $m_W$,
\begin{equation}\label{Z}
\begin{split}
  Z &= \sum_{MN} \frac{\zeta_0^M}{M!} \frac{\tilde\zeta_0^N}{N!}
   \sum_{e_i=\pm1}\sum_{g_j=\pm1}\int\!\prod_{i=1}^M d^2x_i
   \prod_{j=1}^N d^2y_j\, \exp\left[
  \frac{e^2T}{2\pi}\sum_{i<j}^M e_ie_j \ln(T|\vec x_i{-}\vec x_j|)\right.\\ 
  &\quad\left.
  +\frac{g^2}{2\pi T}\sum_{i<j}^N g_i g_j \ln(T|\vec y_i{-}\vec y_j|)
  +
  2i \sum_{i=1}^M\sum_{j=1}^N e_ig_j \theta(\vec x_i-\vec y_j)
  \right] \,.
\end{split}
\end{equation}
Here $\zeta_0$ and $\tilde\zeta_0$ are the instanton and $W$
fugacities, respectively.  They have dimension 2 and are both
exponentially suppressed, 
\begin{subequations}\label{bare}
\begin{eqnarray}
  \zeta_0 & \sim & T^2e^{-S_0} \,,\\
  \tilde\zeta_0 & \sim & T^2e^{-m_W/T}.
\end{eqnarray}
\end{subequations}
(The pre-exponents will not be needed for our computation of $T_c$.)
We put the subscript ``0'' in the notations because these fugacities
serve as the initial conditions for the renormalization group in
Sec.~\ref{sec:RG}.  The sums over $e_i$ and $g_i$ are restricted to
configurations with zero total electric and magnetic charges.

\section{Effective field theories}
\label{sec:effective}

In this section we (re)derive effective bosonic and fermionic theories
describing a 2d gas of electric and magnetic charges. Our bosonic
Lagrangian is different from the one obtained by the authors of
\cite{DKKT}.

\subsection{Bosonic theory}

By using the result of Appendix A, one can write each term in the
sum~(\ref{Z}) as a multipoint Green's function,
\begin{equation}\label{ZPhiTheta}
  Z = \sum_{MN} \frac{\zeta_0^M}{M!} \frac{\tilde\zeta_0^N}{N!}
   \sum_{e_i=\pm1}\sum_{g_j=\pm1}\int\!
   \prod_{i=1}^M d^2x_i\prod_{j=1}^N d^2y_j\,
  \left\< \mbox{T}\prod_i e^{ie_i\beta\Phi(\x_i)} 
  \prod_j e^{ig_j\tilde\beta\Theta(\y_j)} \right\>\,,
\end{equation}
where we have introduced
\begin{equation}\label{betaT}
  \beta = e\sqrt T\,,\quad \tilde\beta = \frac{g}{\sqrt{T}} \,,
\end{equation}
and the average is taken in the free massless bosonic field theory
with the ultraviolet cutoff at the scale $T$. $\Phi$ is a (scalar)
bosonic field and $\Theta$ is the dual field.

One can now perform the summation over $e_i$ and $g_j$ and over $M$ and
$N$ in \eq{ZPhiTheta}.  The result reads
\begin{equation}
  Z = \left\<T\exp\left[\int\!d^2x\,\Bigl(2\zeta_0\cos\beta\Phi+
  2\tilde\zeta_0\cos\tilde\beta \Theta\Bigr)\right]\right\>\,.
\end{equation}
The right hand side, according to the well-known formula of
perturbation theory, is equal to
\begin{equation}
  Z = \frac{\Tr e^{-HT}}{\Tr e^{-H_0T}}\,,
\end{equation}
where $T$ is the total Euclidean time and
\begin{eqnarray}
  H_0 &=& \int\!dx\,\left[ 
   \frac12(\d_x\Phi)^2+\frac12(\d_x\Theta)^2 \right]\,,\label{H0}\\
  H &=& \int\!dx\,\left[\frac12(\d_x\Phi)^2+\frac12(\d_x\Theta)^2
  + 2\zeta_0\cos\beta\Phi + 2\tilde\zeta_0\cos\tilde\beta\Theta\right]\,.
  \label{bosonic}
\end{eqnarray}
Thus, our problem is reduced to finding the ground state energy of the
Hamiltonian $H$ from \eq{bosonic}, defined with an ultraviolet cutoff
at $T$.

One can construct a (Euclidean) Lagrangian corresponding to the
Hamiltonian of \eq{bosonic} by requiring that the free part of the
Lagrangian gives the same correlation functions as in
Eqs. (\ref{PhiPhi}) and (\ref{PhiTheta}). The result is
\be\label{boslagr1}
{\cal L}_b \, = \, \frac{1}{2} (\d_x\Phi)^2 + \frac12(\d_x\Theta)^2 -
i \d_x\Phi \, \d_\tau\Theta + 2 \zeta_0 \cos\beta\Phi +
2 \tilde\zeta_0 \cos\tilde\beta\Theta,
\ee
where $\Phi$ and $\Theta$ should now be understood as two independent
fields. Note that the Lagrangian in \eq{boslagr1} is different from
the one derived in \cite{DKKT}. The main difference is that our
Lagrangian (\ref{boslagr1}) is written for two independent scalar
fields and the Lagrangian used in \cite{DKKT} employs a scalar field
$\chi$ and its dual field $\tilde \chi$ satisfying the duality
condition
\be
\d_\mu \tilde\chi \, = \, - i \epsilon_{\mu\nu} \d_\nu \chi. \label{dual}
\ee
However, the duality condition (\ref{dual}) implies $\d^2 \chi =
\d^2 \tilde\chi = 0$, that is the dual field in the sense of \eq{dual} 
exists only for harmonic functions. It is therefore difficult to
define Feynman functional integrals over $\chi$ for the partition
function employing the Lagrangian which uses both fields $\chi$ and
$\tilde\chi$ since the integration would have to be restricted to
harmonic functions only.

Defining a vector field
\be\label{vf}
C_\mu \, \equiv \, \d_\mu \Phi - i \epsilon_{\mu\nu} \d_\nu \Theta 
\ee
the Lagrangian of \eq{boslagr1} could be rewritten in a more Lorentz
(rotational)-invariant form
\be\label{boslagr2}
{\cal L}_b \, = \,  \frac12(\d_\mu\Theta)^2 + \frac12 (n_\mu C_\mu)^2 +
2\zeta_0\cos\beta\Phi + 2\tilde\zeta_0\cos\tilde\beta\Theta,
\ee
where $n_\mu$ is a unit vector in a randomly chosen direction in the
$(\tau,x)$ plane. Putting $n_\mu = (0,1)$ one would obtain the
Lagrangian in \eq{boslagr1}.

In the limit of no monopoles we put $\zeta_0 = 0$ in \eq{boslagr2}.
Integrating out the $\Phi$ field yields us the Lagrangian of the
sine-Gordon theory for the 2d gas of $W$ bosons
\be\label{sgw}
{\cal L}_{SG}^W \, = \, \frac12(\d_\mu\Theta)^2 +
2\tilde\zeta_0\cos\tilde\beta\Theta. 
\ee
Similarly putting $\tilde\zeta_0 = 0$ in \eq{boslagr2} would eliminate
$W$ bosons from the theory and would give us a sine-Gordon theory for
monopole gas, similar to \cite{AZ}
\be\label{sgm}
{\cal L}_{SG}^{mon} \, = \, \frac12(\d_\mu\Phi)^2 +
2 \zeta_0 \cos\beta\Phi. 
\ee
The exact solution for the $S$-matrix of the sine-Gordon theory is
known \cite{SG} and the exact expression for the soliton mass has been
found \cite{Zamolodchikov}, allowing for quantitative study of the
limits shown in Eqs. (\ref{sgw}) and (\ref{sgm}) along the lines of
\cite{AZ}. An exact solution of the full theory given by \eq{boslagr2}
has not been found. This will prevent us from making quantitative
predictions about the behavior of the string tension around $T_c$ but
will not interfere with our determination of $T_c$ itself.

Polyakov loops play the role of the order parameter of the deconfining
phase transition \cite{SvetitskyYaffe}.  To compute the correlation
function of two Polyakov loops, one inserts two fundamental charges
into the Coulomb gas~(\ref{Z}).  This means that Polyakov loop $P(\x)$
is mapped to the operator $e^{i\tilde\beta\Theta/2}$ of the effective
theory, so that~\cite{AZ,DKKT}
\begin{equation}
  \< P(\x) P^\dagger(0) \> \sim 
  \left\< \exp\left(\frac i2\tilde\beta\Theta(\x)\right) 
  \exp\left(-\frac i2\tilde\beta\Theta(0)\right)\right\>
\end{equation}
where the average in the right hand side is taken in the theory
described by Eq.~(\ref{bosonic}).

\subsection{Fermionic theory}

It is easy to apply the bosonization rules~\cite{Mandelstam} to map
the bosonic theory~(\ref{bosonic}) onto a fermionic theory.  We
noticed above that without the $\tilde\zeta_0 \cos\tilde\beta\Theta$
term the Lagrangian of \eq{boslagr2} reduces to a sine-Gordon theory
given by \eq{sgm}. As was shown in \cite{Mandelstam} the sine-Gordon
theory is equivalent to the massive Thirring model with Dirac mass
term. Mandelstam's rules \cite{Mandelstam} map $\tilde
\zeta_0 \cos\tilde\beta\Theta$ onto a Majorana mass term.  Thus, the
fermionic effective theory is given by the Lagrangian
\begin{equation}
  {\cal L}_f = \bar\psi\gamma^\mu\d_\mu\psi + \frac G2
  (\bar\psi\gamma^\mu\psi)^2
  + m\bar\psi\psi + \frac{\tilde m}2 (\bar\psi^c\psi +\bar\psi\psi^c)
  \label{Lfermion}
\end{equation}
where $\psi^c=C\bar\psi^T$ is charge-conjugated to $\psi$.  In the
chiral basis
\begin{equation}
  \psi=\left( \begin{array}{c} \psi_1 \\ \psi_2 \end{array}\right)\,,\qquad
  \psi_c = \left( \begin{array}{r} \psi_1^\dagger \\ \
  -\psi_2^\dagger \end{array}\right)\,.
\end{equation}
The four-fermion coupling $G$ is related to the parameters of the
bosonic theory by
\begin{equation}\label{gbeta}
  1+\frac G\pi = \frac{4\pi}{\beta^2} = \frac{\tilde\beta^2}{4\pi}\,,
\end{equation}
while the fermion bare masses are
\begin{equation}\label{fmass}
  m \sim \frac{\zeta_0}T\,,\qquad \tilde m\sim \frac{\tilde\zeta_0}T\,.
\end{equation}
One can check by an explicit calculation that the Lagrangian
(\ref{Lfermion}) gives the partition function of \eq{Z}.

Defining real Majorana fermion fields \cite{DKKT}
\begin{equation}\label{maj}
  P = \frac{\psi+\psi^c}{2}\,,\qquad
  Q = \frac{\psi-\psi^c}{2 \, i}
\end{equation}
we can rewrite the Lagrangian of \eq{Lfermion} as
\be\label{lmaj}
{\cal L}_f \, = \, {\bar P} \gamma_\mu \d_\mu P \, + \,
 {\bar Q} \gamma_\mu \d_\mu Q \, - \, 2 \, G \, {\bar P} P \, 
 {\bar Q} Q  \, + \,  (m + {\tilde
m}) \, {\bar P} P \, + \, (m - {\tilde m}) \, {\bar Q} Q.
\ee
The physical meaning of the Lagrangian in \eq{lmaj} will be clarified
in the next Section.

The bosonic theory (\ref{boslagr1}), and, therefore, the partition
function, is symmetric under duality transformation $\zeta_0
\leftrightarrow \tilde\zeta_0$ and $\beta
\leftrightarrow \tilde\beta$, or, equivalently under 
$\Phi\leftrightarrow\Theta$.  The duality transformation
$\Phi\leftrightarrow\Theta$ corresponds to
$\psi_1\leftrightarrow\psi_1$, $\psi_2\leftrightarrow\psi_2^\dagger$.
This transformation interchanges the Dirac and Majorana mass terms,
and the U(1) and U(1)$_A$ currents $j^\mu=\bar\psi\gamma^\mu\psi$ and
$j^{5\mu}=\bar\psi\gamma^\mu\gamma^5\psi$.  Naively under this
transformation the interaction term changes sign,
$G\leftrightarrow-G$, which is not consistent with
$\beta\leftrightarrow\tilde\beta$ in Eq.~(\ref{gbeta}).  This is
because the symmetry between the Dirac and Majorana mass terms in
Eq.~(\ref{Lfermion}) is broken by the implicit procedure of
regularization: when computing the loop graphs in the
theory~(\ref{Lfermion}) one should use a regularization scheme which
preserves the conservation of $j^\mu$ when $\tilde m=0$, e.g., the
Pauli-Villars scheme with a fermion with large Dirac mass
\cite{ZJ}. Using a different regularization scheme, for instance the
Pauli-Villars scheme with a heavy Majorana fermion, would result in a
relation between fermionic and bosonic couplings different from
\eq{gbeta} \cite{ZJ}.

\section{Renormalization group, mass gap and critical temperature}
\label{sec:RG}

\subsection{Review of RG in sine-Gordon theory}

At any temperature, the correlation length of the Georgi-Glashow model
is equal to the inverse mass gap in the effective
theories~(\ref{boslagr2}) and (\ref{Lfermion}).  The fact that there
are two perturbations on top of the free bosonic theory in
Eq.~(\ref{boslagr2}) is crucial for the existence of a critical point
with infinite correlation length (zero mass gap).  To understand the
mechanism of the disappearance of the mass gap, let us recall how the
gap emerges when there is only {\em one} perturbation.

We suppose, for a moment, that $\tilde\zeta_0=0$.  Then our theory is
reduced to the sine-Gordon (SG) model of \eq{sgm}.  The full spectrum
of the SG theory is known~\cite{SG,Zamolodchikov}, but for our
purposes we only need a method for a parametric estimate of the gap.
The Wilson renormalization group (RG) can be employed to this end.  In
this procedure we decrease the ultraviolet cutoff $\Lambda$ by
successively integrating out the modes with momenta above the cutoff,
but below the value of the cutoff at the previous RG step.  Since the
fugacity $\zeta_0$ will run with $\Lambda$ we will denote the
renormalized fugacity by the function $\zeta(\Lambda)$.

When $\zeta$ is small, the perturbation is small and the RG equation
for $\zeta$ is set solely by the conformal dimension of the operator
$\cos\beta\Phi$ \cite{RG,ZJ}
\begin{equation}\label{RGSG}
  \frac\d{\d\lambda}\,\frac\zeta{\Lambda^2} =
  (2-\Delta)\frac\zeta{\Lambda^2}\,,\qquad
  \lambda = \ln \frac T{\Lambda}\,,\qquad \Delta=\frac{\beta^2}{4\pi}\,,
\end{equation}
which has an obvious solution
\begin{equation}
  \frac\zeta{\Lambda^2} = \frac{\zeta_0}{T^2} 
  \left(\frac T\Lambda\right)^{2-\Delta}\,,\qquad 
  \Delta = \frac{\beta^2}{4\pi}\,.
\end{equation}
We have written the RG equation at the leading order in the
dimensionless coupling $\zeta/\Lambda^2$ which controls the conformal
perturbation theory.  If $\beta^2>8\pi$ the perturbation is irrelevant
since $\zeta/\Lambda^2$ is smaller than $\zeta_0/T^2$ for all $\Lambda
< T$.  In the case of $\beta^2<8\pi$ where the perturbation is
relevant, there exists a scale where $\zeta/\Lambda^2 \sim1$ and the
perturbation is no longer small.  Our RG equation~(\ref{RGSG}) is not
valid beyond this point.  The theory becomes strongly coupled at this
scale and one can expect the emergence of a mass gap.  Thus the mass
gap is of the order
\begin{equation}\label{Lambda1}
  \Lambda\sim T \, \left(\frac{\zeta_0}{T^2}\right)^{1/(2-\Delta)} \sim
  T \, e^{-S_0/(2-\Delta)}
\end{equation}
which agrees parametrically with the result of
~\cite{Zamolodchikov,SG}.  The nature of the RG method does not allow
the reproduction of the pre-exponents in Ref.~\cite{Zamolodchikov},
for which one actually needs the exact expression for the $S$-matrix
of the theory.

\subsection{Renormalization group and criterion for $T_c$}

Now we consider our case, where both perturbations are present [see
Eqs. (\ref{bosonic}) and (\ref{boslagr2})].  We now have two RG
equations
\begin{subequations}
\begin{eqnarray}
  \frac\d{\d\lambda}\,\frac{\zeta}{\Lambda^2} &=&
  (2-\Delta)\frac\zeta{\Lambda^2}\,,\qquad
  \Delta = \frac{e^2T}{4\pi}=\frac{4\pi T}{g^2}\\
  \frac\d{\d\lambda}\,\frac{\tilde\zeta}{\Lambda^2} &=&
  (2-\tilde\Delta)\frac{\tilde\zeta}{\Lambda^2}\,,\qquad
  \tilde\Delta = \frac{g^2}{4\pi T}\,.
\end{eqnarray}
\end{subequations}
with solutions
\begin{equation}
  \frac{\zeta}{\Lambda^2} = \frac{\zeta_0}{T^2}
    \left(\frac T\Lambda\right)^{2-\Delta}\,,\qquad
  \frac{\tilde\zeta}{\Lambda^2} = \frac{\tilde\zeta_0}{T^2}
    \left(\frac T\Lambda\right)^{2-\tilde\Delta}\,.
\end{equation}
Noting that $\Delta\tilde\Delta=1$, one can immediately distinguish
three temperature regimes.
\begin{itemize}
\item[(i)] $T<g^2/(8\pi)$.  In this case $\Delta<\frac12$,
$\tilde\Delta>2$.  The operator $\cos\tilde\beta\Theta$ is irrelevant
and can be ignored.  The infrared properties of the theory is the same
as of the sine-Gordon theory (\ref{sgm}) and hence the theory has a
mass gap \cite{AZ,Zamolodchikov}.
\item[(ii)] $T>g^2/(2\pi)$.  In this case $\Delta>2$,
$\tilde\Delta<\frac12$, the operator $\cos\beta\Phi$ is irrelevant and
the theory is in the massive phase of the SG theory of \eq{sgw}.
\item[(iii)] $g^2/(8\pi)<T<g^2/(2\pi)$.  In this case both
perturbations are relevant. 
\end{itemize}

The critical point should be in the regime (iii) which we now
consider.  If we follow the Wilson RG procedure, then at first, when
$\Lambda \sim T$, both perturbations are small.  As one decreases
$\Lambda$ both perturbations become more and more important.  However,
if at some scale one perturbation is of order one while the other is
still small, then one can conclude that the theory has a mass gap.
Indeed, suppose that at some $\Lambda$, $\zeta/\Lambda^2\sim1$ while
$\tilde\zeta/\Lambda^2\ll1$.  In this case the theory is a SG model
with a very small perturbation $\sim\tilde\zeta$.  Since the SG is
fully gapped, the perturbation theory over $\tilde\zeta/\Lambda^2$ is
non-degenerate and does not destroy the existence of the gap.  The
same happens when $\tilde\zeta/\Lambda^2$ becomes $\sim1$ when
$\zeta/\Lambda^2$ is still small.  Therefore, the theory can be
critical only when $\zeta/\Lambda^2$ and $\tilde\zeta/\Lambda^2$ are
of order one at the same energy scale.

\begin{figure}[h]
\setlength{\unitlength}{1.5mm}
\begin{picture}(70,35)(0,7)
\put(10,10){\vector(1,0){50}}
\put(10,10){\vector(0,1){30}}
\put(20,35){\line(3,-1){30}}
\put(20,35){\line(3,-2){30}}
\put(10,35){\line(1,0){1}}
\put(50,10){\line(0,1){1}}
\put(20,10){\line(0,1){1}}
\put(59,7){$\Lambda$}
\put(49,7){$T_c$}
\put(19,8){$\mu$}
\put(7,34){1}
\put(51,24){$\tilde\zeta/\Lambda^2$}
\put(51,14){$\zeta/\Lambda^2$}
\end{picture}
\caption{The running of fugacities at $T=T_c$.}
\label{fig:running}
\end{figure}
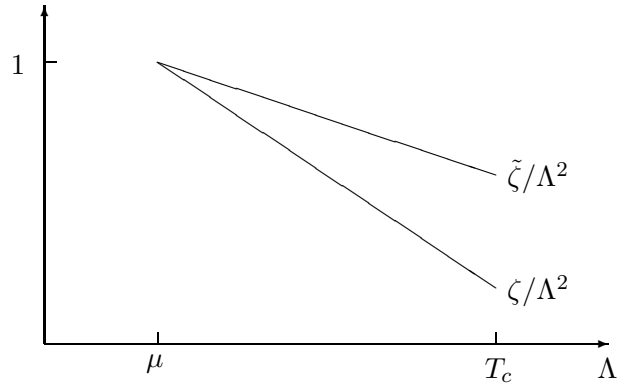

In Fig.~\ref{fig:running} we illustrate what happens at $T_c$.  At the
scale $T=T_c$ we chose the monopole fugacity $\zeta$ to be much
smaller than the $W$ fugacity $\tilde\zeta$. (In principle it can be
the other way around as well.)  However, $\zeta$ runs faster than
$\tilde\zeta$, so that at some scale $\mu$ the two fugacities become
equal and are of the order $\mu^2$.  In other words, the perturbative
expansion over the two fugacities breaks down at the same scale $\mu$
where $\zeta/\mu^2 \sim \tilde\zeta/\mu^2 \sim 1$.  We can guess that
when this condition is satisfied the gap disappears.

If our guess is correct, then $T_c$ can be found be requiring the
existence of a scale $\mu$ where
\be\label{tc}
\frac{\zeta (\mu)}{\mu^2} \, \sim \, \frac{\tilde\zeta (\mu)}{\mu^2} 
\, \sim \, 1.
\ee
Using \eq{bare} in \eq{tc} yields
\begin{equation}\label{Tc_criterion}
  e^{-S_0} \left( \frac T\mu\right)^{2-\beta^2/4\pi} \sim e^{-M_W/T}
  \left( \frac T\mu\right)^{2-\tilde\beta^2/4\pi} \sim 1.
\end{equation}
Solving this equation, taking into account Eqs.~(\ref{Sinst}) and
(\ref{betaT}), we then find our final result for $T_c$,
\begin{equation}\nonumber
  T_c = \frac{g^2}{4\pi} \frac{\epsilon+2}{2\epsilon+1}\,,
\end{equation}
and an estimate for the non-perturbative scale $\mu$ at the critical
temperature
\begin{equation}\label{muTc}
  \mu \sim T \exp\left(-\frac{2\epsilon+1}3\frac{4\pi
  v}g\right)\,,\qquad T \approx T_c.
\end{equation}

We notice, however, that the RG procedure above does not prove that
the temperature found above is the critical temperature: it merely
says that if the theory becomes critical at some temperature, then
this temperature should be given by Eq.~(\ref{Tc}).

To argue that the mass gap vanishes at $T_c$ the fermionic Lagrangian
of \eq{lmaj} becomes useful.  This can be seen most clearly in the
limit of the light Higgs, $m_H\ll m_W$~\cite{DKKT}. (We still have
$m_H \gg g^2$, or, equivalently, $m_H/m_W \gg g/v$, so that one loop
corrections to the monopole action (\ref{Sinst}) would be finite and
\eq{Sinst} would still be valid \cite{KS}.) Then $T_c=g^2/(4\pi)$ with
$\beta^2=\tilde\beta^2=4\pi$, which leads to $G=0$ [see \eq{gbeta}].
The Lagrangian~(\ref{lmaj}) becomes that of a theory with two free
Majorana fermions with masses $m_P=m+\tilde m$ and $m_Q=|m-\tilde
m|$. Using \eq{fmass} with the critical temperature condition \eq{tc}
we may argue that at $T_c$ one may get $m=\tilde m$. When $m=\tilde m$
we obtain $m_P = 2 m$ and $m_Q = 0$.  One of the Majorana fermions is
massless, and the other one is massive.  The existence of the massless
Majorana fermion near the phase transition is in agreement with the
Ising nature of the phase transition~\cite{DKKT}.  Notice that,
strictly speaking, the RG equations can give us $m_P\sim\mu$, but
cannot show that $m_Q=0$.  In the bosonic language, there is no mass
gap when $\beta^2=\tilde\beta^2=4\pi$ and
$\zeta_0=\tilde\zeta_0$~\cite{Ogilvie}.

When the ratio $m_H/m_W$ is large, the fermion coupling $G$ is also
large.  While we do not know the exact mass spectrum of the
theory~(\ref{lmaj}), one may guess that qualitatively the picture does
not change.  This means that near $T_c$ the theory contains two
Majorana fermions: a very light one with some mass $m_Q$, and another
one with mass $m_P\sim\mu\gg m_-$.  Because $G\neq0$ the two fermions
interact with each other.  However, at energy much below $\mu$, the
heavier fermion $P$ decouples, leaving the dynamics to be completely
determined by the lighter fermion $Q$.  Since in 2d there is no
relevant or marginal self-coupling of one Majorana fermion (e.g.,
$(\bar Q Q)^2=0$), the dynamics below $\mu$ is determined by one free
Majorana fermion, as in the limit $m_H/m_W\to0$.

\subsection{Correlation length, critical region, etc.}

It is instructive to find the behavior of the correlator of two
Polyakov loops when $T$ is very close to $T_c$.  This correlator has
various behaviors at different length scales:
\begin{itemize}
\item[(i)] At very large distances $|\x|\gg M^{-1}$, where $M$ is the
mass gap in the theory, 
\be
\<P(\x)P(0)\>\sim e^{-Mx}.
\ee
\item[(ii)] In the intermediate range $\mu^{-1}\ll |\x| \ll M^{-1}$,
where $\mu$ is defined in Eq.~(\ref{muTc}), the correlator behaves
that that of the order parameter in the critical Ising model, so that
\be
\<P(\x)P(0)\>\sim |\x|^{-1/4}.
\ee
\item[(iii)] At distances $T^{-1}\ll |\x|\ll M^{-1}$, the
effective theory is the theory of one massless scalar field, and the
behavior of the Polyakov loop correlator is determined by the
conformal dimension of the operator $e^{i\tilde\beta\Theta/2}$,
\begin{equation}\label{crcont}
  \<P(\x)P(0)\> \sim |\x|^{-(2\epsilon+1)/(2\epsilon+4)}.
\end{equation}
\end{itemize}
It is interesting that this model has Ising-type behavior only at
distances larger than $\mu^{-1}$, which is a length scale
exponentially large compared to the inverse temperature.  At distances
between $T^{-1}$ and $\mu^{-1}$ the correlator of Polyakov loops has a
power-law behavior, but with the critical exponent in \eq{crcont}
which does not have an absolute fixed value and can vary continuously
depending on $m_H/m_W$.

From the discussion above it also follows that the critical region,
i.e., the range of temperatures where the model exhibits the
Ising-like behavior, is very narrow.  For a temperature $T$ to be in
this region, the fugacities $\zeta$ and $\tilde\zeta$ should become
large ($\sim\Lambda^2$) almost at the same scale, where ``almost''
means a mismatch by a factor not much larger than one.  This gives
\begin{equation}
  \frac{| T - T_c |}{T_c} \lesssim \frac gv
\end{equation}
as the critical region.  Deep inside this region, the correlation
length exhibits the Ising behavior with the critical index $\nu=1$:
\be
\xi \sim |T-T_c|^{-1}.  
\ee

Outside the critical region, the correlation length is set by the
energy scale at which one of the perturbations $\cos\beta\Phi$ and
$\cos\tilde\beta\Theta$ becomes large.  Thus,
\begin{equation}\label{corr}
  \xi \sim \left\{ \begin{array}{l} \displaystyle{\frac1T \exp\left(
    \frac{2\pi\epsilon gv}{g^2-2\pi T}\right)}\,,\qquad T<T_c 
  \rule[-.3in]{0in}{.3in}
  \\
  \displaystyle{\frac1T  \exp\left(
    \frac{4\pi gv}{8\pi T-g^2}\right)}
    \,,\qquad T>T_c\,. \end{array} \right.
\end{equation}
One can see from \eq{corr} that as the temperature increases, the
correlation length first increases exponentially from
$m_\gamma^{-1}\sim e^{S_0/2}$ at small $T$ to $\mu^{-1}$ when $T$ is
near the critical region, and then decreases exponentially as $T$
grows past the critical region.  Inside the critical region itself,
$\xi$ in \eq{corr} has the same exponential factor as $\mu^{-1}$ in
\eq{muTc} but the prefactor of the expression for $\xi$ diverges at $T_c$.  
Finally we note that in our model, the correlation length of \eq{corr}
is exponentially large for all $T\ll m_W \sim g v$.

\section{Discussion and Conclusion}
\label{sec:concl}

We have shown that the phase-transition temperature of the
Georgi-Glashow model is given by \eq{Tc}. The critical temperature is
of order $g^2$ and also depends on the ratio $m_H/m_W$ via the
function $\epsilon(m_H/m_W)$ defining the monopole action through
\eq{Sinst}.  In the Georgi-Glashow model $\epsilon$ runs from
$\epsilon(0)=1$ to $\epsilon(\infty)\approx1.787$ \cite{bps,kz}.  If
we formally take $\epsilon \rightarrow 0$ limit in \eq{Tc}, which of
course can never be physically realized in Georgi-Glashow model, we
get $T_c=g^2/(2\pi)$.  This is the value of the critical temperature
suggested in Ref.~\cite{AZ}.  This is not surprising since when
$\epsilon \to 0$ the monopole fugacity becomes $\zeta_0 \sim 1$ while
the $W$ bosons' fugacity is still small $\tilde\zeta_0 \ll 1$, so that
the $W$ bosons are completely outnumbered by the monopoles and the
problem reduces to physics of monopole-only gas considered in
Ref.~\cite{AZ}.  In the opposite unphysical limit of
$\epsilon\to\infty$, the critical temperature in \eq{Tc} becomes $T_c
= g^2/(8\pi)$.  In this limit there is no monopoles and the phase
transition temperature is simply that of the BKT phase transition for
the 2d gas of $W$'s.

In the light Higgs limit $g/v \ll m_H/m_W \ll 1$ corresponding to
$\epsilon = 1$ our critical temperature from \eq{Tc} becomes
$T_c=g^2/(4\pi)$ in agreement with Ref.~\cite{DKKT}. For other values
of $\epsilon$, or, equivalently, for not extremely small values of the
ratio $m_H/m_W$, our critical temperature is different from the one
found in \cite{DKKT}.

The methods used in our paper are almost identical to those of
Ref.~\cite{DKKT}.  The crucial difference is that in Ref.~\cite{DKKT}
the RG equation is run to $\lambda=\ln(T/\Lambda)=\infty$, reaching
what the authors of \cite{DKKT} interpret to be a fixed point at
$\zeta=\tilde\zeta=\infty$.  In contrast, in our paper the RG equation
is used only up to the point where one of the fugacities is of order
one.  Strictly speaking, the RG equations are derived at the leading
order of perturbation theory in $\zeta/\Lambda^2$ and
$\tilde\zeta/\Lambda^2$, and thus cannot be used outside the regime
$\zeta/\Lambda^2,\tilde\zeta/\Lambda^2 \ll 1$.

There exists an intuitive argument leading to our result~(\ref{Tc}).
According to this argument, the phase transition occurs when the
densities of monopoles and $W$ bosons are equal \cite{DKKT}.  Naively
the instanton density is $e^{-S_0}$, but in 2d the instanton action
acquires a logarithmic contribution from distances between $1/T$ and
the mean inter-instanton distance $l_{mon}$.  Therefore $l_{mon}$ can
be obtained from the consistency condition equating the monopole
density obtained from geometric considerations $\sim 1/l_{mon}^2$ to
the same density obtained from the modified monopole action. The
condition reads
\begin{equation}\label{lm}
  \frac{1}{l_{mon}^2} \sim T^2 \exp\left[-S_0 -
  \frac{e^2T}{4\pi}\ln(T l_{mon})\right]\,.
\end{equation}
Analogously, the mean distance between the $W$ bosons $l_W$ is given by
\begin{equation}\label{lW}
  \frac{1}{l_W^2} \sim T^2 \exp\left[-\frac{m_W}T - \frac{g^2}{4\pi
  T} \ln(Tl_W)\right]\,.
\end{equation}
Solving Eqs.~(\ref{lm}) and (\ref{lW}) for $l_{mon}$ and $l_W$, and
requiring $l_m=l_W$, one then obtains Eq.~(\ref{Tc}).\footnote{Almost
the same argument was given in Ref.~\cite{DKKT}, but the logarithmic
terms in the exponents were missing.}

\begin{acknowledgments}

The authors thank Guy Moore, Larry Yaffe, and especially Alex Kovner
for informative discussions. The work of Yu.~K. is supported in part
by the U.S.\ Department of Energy under Grant No.\ DE-FG03-97ER41014 and
by the BSF grant $\#$ 9800276 with Israeli Science Foundation, founded
by the Israeli Academy of Science and Humanities. The work of
D.~T.~S. is supported in part by the U.S. Department of Energy under
Grant No.\ DOE-ER-41132 and by the Alfred P.\ Sloan Foundation.

\end{acknowledgments}

\appendix
\section{Free massless boson in 2d}

We consider a free massless boson in two dimensions,
\begin{equation}\label{SPhi}
  S = \int\!d^2x\, \frac12 (\d_\mu\Phi)^2\,.
\end{equation}
Although Eq.~(\ref{SPhi}) is invariant under O(2) rotations, we can
choose an arbitrary direction to be the Euclidean time axis, and
quantize the theory canonically.  The canonical commutation relation
reads
\begin{equation}
  [\Phi(\tau,x),\, \d_\tau\Phi(\tau, y)] = \delta(x-y)\,,
\end{equation}
and the Hamiltonian is
\begin{equation}\label{HPhi}
  H = \frac12 \int\!dx\, [(-\d_\tau\Phi)^2 + (\d_x\Phi)^2]\,.
\end{equation}
We introduce a field $\Theta$ defined as follows,
\begin{equation}\label{Theta-def}
  \Theta(\tau,x) = i\int\limits_{-\infty}^x\!dy\, \d_\tau\Phi(\tau,y)\,.
\end{equation}
Since $\tau$ is the Euclidean time, $\Theta$ is a Hermitian operator.
By definition, $\d_x\Theta=i\d_\tau\Phi$.  One also finds
$\d_\tau\Theta=-i\d_x\Phi$ by using the free field equation satisfied
by $\Phi$.  Thus, $\d_\mu\Theta=-i\epsilon_{\mu\nu}\d_\nu\Phi$.  We
will say that $\Theta$ is dual to $\Phi$.  The commutation relation
between $\Phi$ and $\Theta$ can also be found from the definition of
$\Theta$~(\ref{Theta-def}),
\begin{equation}
  [\Phi(\tau,x),\, \Theta(\tau,y)] = i\theta(y-x)\,.
\end{equation}
In other words, $\Phi$ and $\Theta$ are mutually non-local.  The
Hamiltonian~(\ref{HPhi}) can be written in a form symmetric under
duality transformation $\Phi\leftrightarrow\Theta$,
\begin{equation}
  H = \frac12 \int\!dx\,\left[(\d_x\Phi)^2 + (\d_x\Theta)^2\right]\,.
\end{equation}

Since~(\ref{SPhi}) is a free field theory, the Euclidean Green's
functions can be easily computed,
\begin{eqnarray}
  \< \mbox{T}\Phi(\x)\Phi(0)\> &=& \< \mbox{T}\Theta(\x)\Theta(0)\> =
  -\frac1{2\pi} \ln(m|\x|) \label{PhiPhi}\\
  \<\mbox{T}\Phi(\x)\Theta(0)\> &=& \frac i{2\pi}\, {\rm sgn}(\tau)
  \arccos\frac x{\sqrt{\tau^2+x^2}} = \frac i{2\pi}\theta(\vec x)
  \label{PhiTheta}
\end{eqnarray}
where $\x=(\tau,x)$ and $m$ is a small infrared regulator.  In
Eq.~(\ref{PhiTheta}) the value of $\arccos$ is taken to be in
$(0,\pi)$.  The function $\theta(\vec x)$ is the angle between the
vector $\vec x=(\tau,x)$ and the $x$ axis, and runs between $-\pi$ to
$\pi$.  The discontinuity of (\ref{PhiTheta}) at $\tau=0$, $x<0$ is
due to the Euclidean time ordering,
\begin{equation} 
  \<\mbox{T}\Phi(\epsilon,x)\Theta(0,0)\>-\<\mbox{T}\Phi(-\epsilon,x)
\Theta(0,0)\>=
  \<[\Phi(0,x),\, \Theta(0,0)]\> = i\theta(-x),
\end{equation}
with $\theta$ here being the $\theta$-function.  The presence of a cut
along the spatial axis means that the correlator~(\ref{PhiTheta}) is
not invariant under O(2) rotations in the Euclidean space-time.  This
is not surprising, since $\Phi$ and $\Theta$ are not mutually local.

The following correlator can now be easily computed
\begin{equation}\label{Coulomb-corr}
\begin{split}
  \left \< \mbox{T}\prod_{i=1}^M e^{i\beta_i\Phi(\vec x_i)}
  \prod_{j=1}^N e^{i\tilde\beta_j\Theta(\vec y_j)}\right\> &=
  \left(\frac\Lambda m\right)^{-((\sum_i\beta_i)^2
  + (\sum_j\tilde\beta_j)^2)/4\pi}
  \exp\left[\sum_{i<j}^M \frac{\beta_i\beta_j}{2\pi}
  \ln(\Lambda|\vec x_i-\vec x_j|)\right.\\
  &\quad+\left.
  \sum_{i<j}^N \frac{\tilde\beta_i\tilde\beta_j}{2\pi}
  \ln(\Lambda |\vec y_i-\vec y_j|) -
  i\sum_{i,j}^{M,N}\frac{\beta_i\tilde\beta_j}{2\pi}\theta(\x_i-\y_j) 
  \right]
\end{split}
\end{equation}
where $\Lambda$ is the ultraviolet cutoff.  The $m$ dependence tells
us that the correlation function vanishes in the limit $m\to0$ unless
$\sum_i\beta_i=\sum_j\tilde\beta_j=0$ (charge neutrality).  The
dependence of the correlator on the ultraviolet cutoff $\Lambda$ is
$\Lambda^{-\Delta}$, where
\begin{equation}
  \Delta = \sum_i \Delta_i + \sum_j \tilde\Delta_j \equiv
  \sum_i \frac{\beta_i^2}{4\pi} 
  + \sum_j \frac{\tilde\beta_j^2}{4\pi}
\end{equation}
is the sum of the conformal dimensions of the operators in the
correlator (\ref{Coulomb-corr}).  The correlator~(\ref{Coulomb-corr})
does not have branch cuts if all products $\beta_i\tilde\beta_j$ are
multiple of $2\pi$; in this case, the result does not depend on the
choice of the time axis.  This is due to the fact that the operators
in the correlator~(\ref{Coulomb-corr}) are mutually local.  For the
correlator in \eq{ZPhiTheta} $\beta\tilde\beta = 4 \pi$ and the
partition function is indeed independent of the choice of the time
axis.

\end{document}